\documentclass[12pt,natbib,authoryear]{elsarticle}
\usepackage{amsmath}
\usepackage{subfigure}
\usepackage{float}
\usepackage{graphicx}
\usepackage{fullpage}
\newcommand{\etal}{\textit{et al}}
\newcommand{\bfd}{\mathbf{d}}
\newcommand{\bfe}{{\mathbf e}}
\newcommand{\bfm}{\mathbf{m}}
\newcommand{\bfr}{\tilde{\mathbf{r}}}
\newcommand{\bfu}{\mathbf{u}}
\newcommand{\Wh}{W_{\mathrm{hard}}}
\newcommand{\Wdepth}{W_{\mathrm{depth}}}
\newcommand{\Wd}{W_{\mathbf d}}
\newcommand{\We}{W_{\mathrm{ e}}}
\newcommand{\argmin}[1]{\textnormal{arg} \min_{#1}}
\newtheorem{algorithm}{Algorithm}
\begin{document}
\journal{Journal of Geophysics and Engineering}
\begin{frontmatter}
\title {Automatic estimation of the regularization parameter in  2-D focusing gravity inversion: an application to the Safo manganese mine in northwest of Iran}
\author[label1]{Saeed Vatankhah}
\ead{svatan@ut.ac.ir}
\author[label1]{Vahid E Ardestani}
\ead{ebrahim@ut.ac.ir}
\author[label2]{ Rosemary A Renaut\corref{cor1}}
\ead{renaut@asu.edu}
\date{\today}
\cortext[cor1]{Corresponding Author: Rosemary Renaut, 480 965 3795 }
\address[label1]{Institute of Geophysics, University of  Tehran,Tehran, Iran} 
\address[label2]{School of Mathematical and Statistical Sciences, Arizona State University, Tempe, AZ 85287-1804, USA}
 
\begin{abstract}We investigate the use of Tikhonov regularization with the minimum support  stabilizer for  underdetermined  2-D inversion of  gravity data. This stabilizer produces models with non-smooth properties which is useful for identifying geologic structures with sharp boundaries. A very important aspect of using Tikhonov regularization is the choice of the regularization parameter that controls the trade off between the data fidelity and the stabilizing functional. The L-curve and generalized cross validation techniques, which  only require the relative sizes of the uncertainties in the observations are considered. Both criteria are applied in an iterative process for which at each iteration a
value for regularization parameter is estimated. Suitable values for the regularization parameter are  successfully determined in both cases for synthetic but practically relevant examples.  Whenever the geologic situation permits, it is easier and more efficient to model the subsurface with a 2-D algorithm, rather than to apply a full 3-D approach. Then, because the problem is not large it is appropriate to use the generalized singular value decomposition  for solving the problem efficiently. The method is applied on a profile of gravity data acquired over the Safo mining camp in Maku-Iran, which is well known for manganese ores. The presented results demonstrate success in reconstructing the geometry and density distribution of the subsurface source.

\end{abstract}
 \begin{keyword}
 Gravity inversion, manganese exploration, regularization parameter, L-curve criterion, generalized cross validation, generalized singular value decomposition

\PACS 93.30.Bz, \sep 93.85.Hj
\end{keyword}
\end{frontmatter}

\section{Introduction}
Gravity inversion reconstructs models of subsurface density distribution using measured data on the surface. There are two important ambiguities in the inversion of gravity data. Theoretical ambiguity is caused by the nature of gravity; many equivalent sources in the subsurface can produce the same data at the surface.  Algebraic ambiguity occurs when  parameterization of the problem creates an underdetermined situation with more unknowns than observations. Then,  there is no unique density distribution which satisfies the observed data. Further, the measurement process at the Earth's surface is necessarily error-contaminated and such errors can introduce arbitrarily large changes in the reconstructed solutions;  namely the solutions are sensitive to errors in the measurements. Thus, the inversion of gravity data with under sampling is a typical example of an ill-posed problem  that requires the inclusion of a priori information in order to find a feasible reconstruction.

Tikhonov regularization is a well-known and well-studied method for stabilizing the solutions of ill-posed problems, \citep[e.g.]{Hansen:98,Vogel:02}. The objective function of the Tikhonov formulation includes a data fidelity (misfit term), and a stabilizing term that controls the growth of the solution with respect to a chosen weighted norm. The choice of the weighting for the regularization term impacts the properties of the solution. For example, a smoothing stabilizer which employs the first or second derivative of the model parameters, such as used  in   \citep[e.g.]{Li,Litwo} produces smooth images of the subsurface density distribution. There are, however, situations in which the potential field sources are localized and have material properties that vary over relatively short distances.  Then, a regularization that does not not penalize sharp boundaries should be used.  \citet{Last} presented a compactness criteria for gravity inversion that seeks to minimize the volume of the causative body. This concept was developed by introducing minimum support (MS) and minimum gradient support (MGS) stabilizers \cite{Port}, \cite{Zhd}, which are applied iteratively, generating  repeatedly updated weighted quadratic stabilizers. 

In any regularization method, the trade off between the data fit and the regularization term  is controlled by a regularization parameter. Methods to find this regularization parameter, called parameter-choice methods, can be divided into two classes \cite{Hansen:98}: (i) those that are based on knowledge of, or a good estimate of, the error in the observations, such as Morozov's discrepancy principle (MDP), and (ii) those that, in contrast, seek to extract such information from the observations, such as the L-curve or generalized cross-validation (GCV) methods. The use of the MDP is dominant in papers related to potential field inversion,\citep[e.g.]{Li,Litwo},  and the original paper for focusing inversion \cite{Port}. In many practical applications, little knowledge about the noise or error in the data measurements is available. The MDP then reduces to a trial and error procedure for finding the optimal regularization parameter \cite{Li3}. Here we discuss the use of the  L-curve and GCV methods for use in focusing inversion of gravity data in situations in which there is information about the  relative magnitudes of the standard deviations across the measured data \cite{Far:04}. 
 Due to the iterative nature of the algorithm, the regularization parameter is determined each iteration. 

Depending on the type of problem to be tackled, gravity inversion can be carried out either in two or three dimensions (2-D or 3-D). 2-D methods are suitable for the recovery of geologic structures such as faults, dikes and rift zones for which the length of the source body in one direction is much longer than its extension in other directions. Then, it may be possible to consider the gravitational sources as completely invariant in the direction parallel to the length direction. Additionally, 2-D sources are both easier to conceptualize and  model than their 3-D counterparts, \cite{Blakely}.  

The outline of this paper is as follows. In Section~\ref{sec2} we review the derivation of the  analytic calculation of the gravity anomaly derived from a 2-D cell model  and then present an overview of  numerical methods for focusing inversion. The  use, and rationale for the use, of the generalized singular value decomposition,  \cite{PaigeSau1}, for the solution  is given  in Section~\ref{gsvdsec}.  The L-curve and GCV  parameter-choice methods are  discussed in  Section~\ref{regparam}.  Results for synthetic data are illustrated in Section~\ref{results}.   The approach is applied on a profile of gravity data acquired from the Safo mining camp in Maku-Iran in  Section~\ref{realresults}.  Future directions and conclusions follow in Section~\ref{sec:conc}.
\section{Gravity modeling}\label{sec2}
\subsection{The theoretical model}
\begin{figure}[htb]
\begin{center}{\includegraphics[width=.8\textwidth]{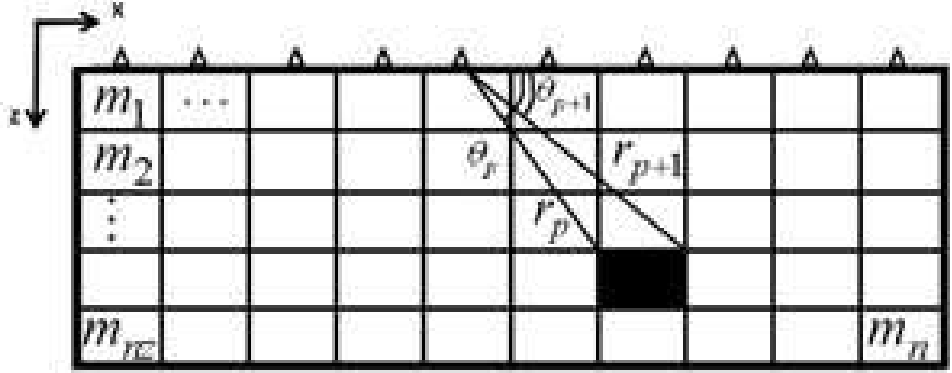}}
\caption{\label{fig2}The cross-section of the subsurface under the gravity profile. Gravity stations are located at the centers of the blocks at the ground surface, indicated by the $\Delta$ symbols. The cells are square and their dimensions are equal to the distances between two observation points. Each cell extends as an infinitely long prism in the invariant $y-$direction.}
\end{center}\end{figure}

A simple 2-D model is obtained by dividing the subsurface under the survey area in to a large number of infinitely long horizontal prisms in the invariant $y-$direction,  with variations in densities  only assumed for the $x$ and $z$ directions. The cross-section of the subsurface under the gravity profile for the model is shown in Figure~\ref{fig2} in which the cells have square cross section and unknown densities.  The dimensions of the cells 
 are equal to the distances between two observation points and the unknown density is considered to be constant for each block.   This type of parameterization, originally used by \cite{Last}, could indeed be improved. Yet, increasing the resolution of the models, and hence the number of parameters, by dividing the subsurface into smaller cells, makes the problem more ill-posed. Here the unknown density is considered to be constant for each block and the data and model parameters are linearly related.
The vertical component of the gravitational attraction $g_i$  of a two-dimensional body at the origin using Cartesian coordinates is given by, \cite{Blakely},
\begin{align}
g_i&= 2 \Gamma \rho \int\int \frac{z'dx'dz'}{{x'}^2+{z'}^2}.
\end{align}
Here  $\Gamma$ is the universal gravitational constant and the density $\rho$   is assumed to be constant within the body. A solution of this integral for an $\ell$-sided polygon is given by, \cite{Blakely},
\begin{align}\label{polygon}
\frac{g_i}{\rho_j}&=2 \Gamma \sum_{p=1}^\ell \frac{\nu_p}{1+\psi_p^2}\left( \log \frac{r_{p+1}}{r_p} - \psi_p (\theta_{p+1}-\theta_p)\right),
\end{align}
where $\psi_p=(x_{p+1}-x_p)/(z_{p+1}-z_p)$ and $\nu_p=x_p-\psi_pz_p$.  The variables $r_p$,  $r_{p+1}$, $\theta_p$, and $\theta_{p+1}$  are as  displayed for the upper side of a square block in Figure~\ref{fig2}.                                                                                               
The term on the right-hand side of  \eqref{polygon}, which quantifies the contribution to the $i$th datum of a unit density in the $j$th cell, is denoted by the kernel  weight $G_{ij}$, and  is valid only at station $i$  for  cell $j$. The total response  for  station $i$  is obtained   by summing over all cells giving 
\begin{align}
g_i &= \sum_{j=1}^{n} G_{ij} \rho_j, \quad i=1,\dots, m, \quad m \le n,
\end{align}
which leads to the matrix equation 
\begin{align}\label{matrix}
\bfd = G \bfm +\bfe,
\end{align}
where we have used the standard notation that vector $\bfd$ is the set of measurements given by the $g_i$, $\bfm$ is the vector of unknown model parameters, here the densities $\rho_j$, and $\bfe$ represents the error in the measurements.  The purpose of the gravity inverse problem is to find a geophysically
plausible density model that reproduces $\bfd$.
\subsection{Numerical approaches for focusing inversion}\label{numerics}
The conventional method for solving ill-posed inverse problems as described by  \eqref{matrix} is based on minimization of the  parametric functional
\begin{align}\label{Tik}
P^{\alpha}(\bfm)= \phi(\bfd) +\alpha^2 S(\bfm).
\end{align}
Here  $\phi(\bfd)$  measures the data fidelity, which 
is usually measured by the weighted discrepancy 
\begin{align}
 \phi(\bfd) = \| \Wd (\bfd -\bfd_{\mathrm{obs}})\|^2_2,
\end{align}
where $\bfd=G\bfm$  is the vector of predicted data,   $\bfd_{\mathrm{obs}}$ contains the observations and $\Wd$ is a data weighting matrix. 
Under the assumption that the noise $\bfe$ is Gaussian and uncorrelated,   $\Wd=\mathrm{diag}(1/\sigma_1, \dots, 1/\sigma_m)$ where   $\sigma_i$ is the standard deviation of the noise in the $i$th datum.  Following  \citet{Far:04},  rather than always assuming that the absolute magnitudes of the error are known, we assume that
  the relative magnitudes of the error can be estimated. Then for an unknown $\sigma_0$ each $\sigma_i$ is reexpressed as $\sigma_i=\sigma_0\tilde{\sigma}_i$ and $\Wd=\mathrm{diag}(1/\tilde{\sigma}_1, \dots, 1/\tilde{\sigma}_m)$. In \eqref{Tik} $S(\bfm)$ is a stabilizing regularization functional and  $\alpha$ is a regularization parameter. 
 
Different choices are possible for   $S(\bfm)$.  Here we use the  minimum support (MS) stabilizer introduced in \cite{Last} and used in \cite{Port}. This stabilizer generates a compact image of a geophysical model with sharp boundaries and,  following \cite{Zhd}, is of the form: 
\begin{align}\label{SF}
S_{\mathrm{MS}}(\bfm) & = (\bfm-\bfm_{\mathrm{apr}})^T \left((\hat{\bfm}-\hat{\bfm}_{\mathrm{apr}})^2+\epsilon^2I)\right)^{-1}(\bfm-\bfm_{\mathrm{apr}})\\
& =(\bfm-\bfm_{\mathrm{apr}})^T\We^2(\bfm-\bfm_{\mathrm{apr}}),\quad \We  = \left((\hat{\bfm}-\hat{\bfm}_{\mathrm{apr}})^2+\epsilon^2I\right)^{-1/2}, \label{We}
\end{align}
where $\hat{\bfm}$ and  $\hat{\bfm}_{\mathrm{apr}}$ are  diagonal matrices of the current model parameters $\bfm$ and an estimate of the model parameters  $\bfm_{\mathrm{apr}}$. If good prior knowledge of the properties of the subsurface  distribution exists, a full model of the expected physical properties can be used for   $\bfm_{\mathrm{apr}}$, otherwise it is often set to $0$.  In $\We$, $\epsilon\ge 0$ is a focusing parameter that is introduced to provide stability as   $\bfm \rightarrow \bfm_{\mathrm{apr}}$ component wise. Small values for $\epsilon$ lead to compact models but also increase the instability in the solution. For large $\epsilon$ the image will not be focused.  In general we are interested in the case where $\epsilon\rightarrow 0$.  A trade-off curve method can be used to select $\epsilon$,  
\cite{ZhdTol,Ajo,Vatan}.  

It is well known that in potential data inversion the reconstructed models tend to concentrate near the surface.  The depth weighting matrix, $\Wdepth=1/(z_j+\zeta)^\beta$   introduced in \cite{Li,Litwo,Pilk}  can then be incorporated into the stabilizer term. Here $z_j$ is the mean depth of cell $j$ and $\zeta>0$ is a small number imposed to avoid singularity at the surface.  The choice for $\beta$ is important.  Small $\beta$   provide a shallow reconstruction for the solution and large values concentrate the solution at depth.  For all inversions considered here $\beta=0.6$ is selected. The hard constraint matrix $\Wh$ is also important for the inversion process. If  field operation geological and geophysical information are able to provide the value of the   density of  cell $j$ this information should be included in $\bfm_{\mathrm{apr}}$.  Then,  $\Wh$ is the diagonal identity matrix  but with $(\Wh)_{jj}=100$ for   those diagonal entries $j$ for which information on the density is known. 

Combining the diagonal weighting matrices, $\Wh$, $\We$ and $\Wdepth$,    \eqref{Tik} is replaced by 
\begin{align}\label{model}
P^{\alpha}(\bfm)=\|\Wd (G \bfm -\bfd_{\mathrm{obs}})\|^2 + \alpha^2 \|D(\bfm-\bfm_{\mathrm{apr}})\|^2, \quad D=\We \Wh \Wdepth.
\end{align}
To obtain   solution $\bfm :=\argmin{\bfm} {P^{\alpha}(\bfm)}$   linear transformation of the original model parameters is introduced via 
$\bfm(\alpha)=\bfm-\bfm_{\mathrm{apr}}$.  Then $\bfm(\alpha)$ solves the normal equations 
\begin{align}\label{normal}
(G^T\Wd^2G+ \alpha^2D^TD) \bfm(\alpha) &= G^T \Wd^2 (\bfd_{\mathrm{obs}}-G\bfm_{\mathrm{apr}}) \quad \mathrm{providing}\\
\bfm & = \bfm_{\mathrm{apr}}+\bfm(\alpha).
\end{align}
When \eqref{model} is solved iteratively due to the dependence of   $\We$ on $k$, we use $\bfm_{\mathrm{apr}}=\bfm^{(k-1)}$,  $\bfm^{(0)}=0$ and $D^{(k)} =\We^{(k)}\Wh \Wdepth$ with
\begin{align}\label{Wk}
\We^{(k)}=\left((\hat{\bfm}^{(k-1)}-\hat{\bfm}^{(k-2)})^2+\epsilon^2 I\right)^{-1/2} \quad k>1, \quad \We^{(1)}=I.
\end{align}
Then $\bfm^{(k)}$ is obtained from the iteratively regularized equation, with updated regularization parameter $\alpha^{(k)}$,
\begin{align} \label{normalk}
(G^T\Wd^2G+ (\alpha^{(k)})^2(D^{(k)})^TD^{(k)}) \bfm(\alpha^{(k)}) &= G^T \Wd^2 (\bfd_{\mathrm{obs}}-G\bfm^{(k-1)})\quad \mathrm{yielding}\\
\bfm^{(k)} &= \bfm^{(k-1)}+ \bfm(\alpha^{(k)}). \label{update}
\end{align} 
This technique in which the weighting in $\We$ is frozen at each iteration, creating the possibility to solve using the standard Tikhonov update, was introduced in the context of focusing inversion  in \cite{Zhd}. Because the MS stabilizer tends to produce the smallest possible anomalous domain we follow the approach of \cite{Port,BoCh:01}  to produce a reliable image of the subsurface when using focusing inversion. Based on geologic information, upper and lower bounds, $\bfm_{\mathrm{min}} \le \bfm_j \le \bfm_{\mathrm{max}}$,  can be determined for the model parameters. If during the iterative process a given density value falls outside the bounds,  projection is employed to force the value back to the exceeded value, and a hard constraint is imposed at that cell via $(\Wh)_{jj}=100$.

\subsubsection{Numerical Solution by the Generalized Singular Value Decomposition}\label{gsvdsec}
We now discuss the numerical procedure for finding the solution to \eqref{normalk}. For large scale problems, iterative methods such as conjugate gradients, or other Krylov methods should be  employed to find $\bfm(\alpha^{(k)})$, \citep[e.g.]{Hansen:98}. For small scale problems it is feasible to use the singular value decomposition (SVD) for the matrix $\tilde{G}=\Wd G$, when matrix $D$ is the identity. Otherwise the generalized singular value decomposition (GSVD), \cite{PaigeSau1}, is needed. But again is effective to use for this problem because it facilitates efficient determination of the regularization parameter. We assume $\tilde{G} \in \mathcal{R}^{m\times n}$, $m<n$, $D\in \mathcal{R}^{n \times n}$ and  $\mathcal{N}(\tilde{G}) \cap \mathcal{N}(\alpha  D) = 0$, where $\mathcal{N}(\tilde{G})$ is the null space of matrix $\tilde{G}$. Then there exist orthogonal matrices $U \in \mathcal{R}^{m\times m}$, $V \in \mathcal{R}^{n \times n}$ and a nonsingular matrix $X \in \mathcal{R}^{n \times n}$ such that  $\tilde{G}=U \Lambda X^T$, $D=V M X^T$  where $\Lambda$ of size ${m \times n}$ is zero except for  entries $0< \Lambda_{1,q+1} \le \dots \Lambda_{m,n} <1$ with $q=n-m$, $M$ is diagonal of size $n\times n$ with entries $1=M_{1,1} =\dots M_{q,q}> M_{q+1,q+1}\ge M_{2,2} \ge \dots M_{n,n}>0$. The generalized singular values of the matrix pair $\tilde{G}$, $D$ are $\gamma_i=\lambda_i/\mu_i$, where $\gamma_1=\dots =\gamma_{q}=0<\gamma_{q+1}\le \dots \le \gamma_n$, and $\Lambda^T\Lambda =\mathrm{diag}(0,\dots 0,\lambda^2_{q+1},\dots, \lambda_n^2)$, $M^T M= \mathrm{diag}(1,\dots,1,\mu_{q+1}^2,\dots,\mu_n^2)$, and $\lambda_i^2+\mu_i^2=1$, $\forall i=1:n$, i.e. $M^T  M + \Lambda^T \Lambda=I_n$. 

Using the GSVD,  introducing $\bfu_i$ as the $i$th column of matrix $U$ and $\bfr^{(k)}=\Wd(\bfd_{\mathrm{obs}}-G\bfm^{(k-1)})$, we may immediately write the solution of \eqref{normalk} as 
\begin{align}\label{gsvdsoln}
&\bfm(\alpha^{(k)}) = \sum_{i=q+1}^{n} \frac{\gamma^2_i}{\gamma^2_i+(\alpha^{(k)})^2} \frac{\bfu^T_{i-q}\bfr^{(k)}}{\lambda_i} (X^T)^{-1}_{i}, \\
&\bfm^{(k)}  = \bfm^{(k-1)}+\sum_{i=q+1}^{n} f_i \frac{\bfu^T_{i-q}\bfr^{(k)}}{\lambda_i} (X^T)^{-1}_{i}, \, f_i=\frac{\gamma^2_i}{\gamma^2_i+(\alpha^{(k)})^2}, \,q<i\le n, \, f_i=0, \,1\le i\le q,  \label{filtersoln}
\end{align}
where $(X^T)^{-1}_{i}$ is the $i$th column of the inverse of the matrix $X^T$ and $f_i$ are the filter factors. Therefore the algorithm proceeds by first updating the matrix $\We$ at step $k$ using \eqref{Wk},  calculating the GSVD for the matrix pair $[\tilde{G}, D^{(k)}]$ and then updating $\bfm^{(k)}$ using \eqref{filtersoln} which depends on $\alpha^{(k)}$. 

Three   criteria are used to terminate  the  iterative procedure. Following \citet{Far:04} the iteration is seen to have converged and is thus terminated when either (i) a sufficient decrease in the functional is observed, $P^{\alpha^{(k-1)}}-P^{\alpha^{(k)}}< \tau(1+P^{\alpha^{(k)}})$, or (ii) the change in the density satisfies $\|\bfm^{(k-1}-\bfm^{(k)} \| < \sqrt{\tau} (1 + \|\bfm^{(k)}\|)$. If neither of these conditions is satisfied by an upper limit on the number of iterations, the procedure is terminated without convergence as measured in this manner. The parameter $\tau$ is taken as $\tau=.01$ for the inversions considered here.

The remaining issue is the determination of  the regularization parameter at each step of the iteration. As noted, when a priori information in the form of the standard deviations on the noise in the data is available, the MDP can be used to find $\alpha$. 
Here, in the absence of the exact information on the error in the data we investigate the use of the L-curve and GCV methods to find $\alpha^{(k)}$ for which the formulation using the GSVD is advantageous.

\subsection{Regularization Parameter Estimation}\label{regparam}
\subsubsection{The L-curve}\label{Lc}
The L-curve approach developed by \cite{Hansen:92,Hansen:98} for linear inverse problems is a robust criterion for determining the regularization parameter. It is  based on the trade-off between the norm of the regularized solution and the norm of the corresponding fidelity term residual as the regularization parameter varies. According to \cite{Hansen:92,Hansen:98}  when these two norms are plotted on a log-log scale, the curve has an L shape with an obvious corner. This corner separates the flat and vertical parts of the curve where the solution is dominated by regularization errors and perturbation errors, respectively. Picking $\alpha_{\mathrm{opt}}$   as the $\alpha$ responsible for the corner point  gives the optimal trade off between the two terms, and the corresponding model is selected as the optimal solution. For $\alpha>\alpha_{\mathrm{opt}}$  the regularized solution does not change dramatically, while the residual does. In contrast, for $\alpha<\alpha_{\mathrm{opt}}$    the regularized solution increases rapidly with little decrease in the residual. Because of the relation of $\alpha_{\mathrm{opt}}$ with the shape of the curve,  \citet{Hansen:98} recommends   estimating $\alpha_{\mathrm{opt}}$ by finding the maximum of the local curvature in the neighborhood of the dominant corner of the plot. Although the L-curve technique can be robust for problems generating well-defined corners, it may not work so well in other cases. For an underdetermined problem the recovered model can change more slowly with the degree of regularization, \cite{Li3}, and the L-curve is thus smoother. This makes it difficult to find maximum point of  curvature of the curve. On the other hand, given the solution of the regularized problem in terms of the GSVD, as in \eqref{gsvdsoln}, finding the L-curve is relatively efficient and is thus one of the conventional ways to estimate $\alpha_{\mathrm{opt}}$. 

 It was shown by \citet{Far:04} that the L-curve choice for $\lambda$   at early iterations may be too small which may lead to inclusion of  excessive structure in the model that needs to be eventually removed, hence requiring more iterations for the inversion. Hence here we follow the approach suggested by \citet{Far:04} and impose a so-called \textit{cooling} process in which $\alpha^{(k)}$ is given by $\alpha^{(k)}=\max \left(c \, \alpha^{(k-1)}, \alpha^*\right)$ where $0.01\le c \le 0.5$ and $\alpha^*$ is the point of maximum curvature of the L-curve.  Moreover, choosing a relatively large value for $\alpha^{(1)}$ improves the performance of the algorithm. We use $\alpha^{(1)}=\max(\gamma_i)/\mathrm{mean}(\gamma_i)$  and $c=0.4$ which work well for the presented inversion examples.
 
 \subsubsection{Generalized Cross Validation}\label{GCV}
The major motivation of using the  GCV to find an optimal value for $\alpha$ is that a good value should predict missing data values. Specifically, if an arbitrary measurement is removed from the data set,  then the corresponding regularized solution should be able to predict the missing observation.   The choice of $\alpha$ should be independent of an orthogonal transformation of the data, \cite{Hansen:98}. The GCV functional is given by
\begin{align}\label{gcc}
GCV(\alpha)&= \frac{\|\tilde{G}\bfm(\alpha)-\bfr\|^2}{\mathrm{trace}(I_m-\tilde{G}G(\alpha))^2}=\frac{\|\tilde{G}\bfm^{(k)}-\tilde{\bfd}_{\mathrm{obs}}\|^2}{ (m -\sum_{i=q+1}^n f_i)^2}=\frac{\| \sum_{i=q+1}^n ( 1- f_i) \bfu_{i-q}^T \bfr\|^2}{( m -\sum_{i=q+1}^n f_i)^2},
\end{align}                                                             
where the final expressions follow immediately from the GSVD, with $ {G}(\alpha) =(\tilde{G}^T\tilde{G}+\alpha^2 D^TD)^{-1}\tilde{G}^T$. Here the  numerator is the squared residual norm and the denominator is effectively the square of the number of degrees of freedom,  \cite{Hansen:98}. It can be shown that the value of   $\alpha$ which minimizes the expected value of the  GCV function is near the minimizer of the expected value of the predictive mean-square error, $\|G\bfm(\alpha)-\bfd_{\mathrm{exact}}\|^2$,  \cite{Hansen:98}.  Hence, finding the minimum for $GCV(\alpha)$ should lead to a reasonable estimate for $\alpha$. We have seen in our experiments that when the GCV does not fail, in which case it produces a very small $\alpha$,  it usually leads to $\alpha$ which is slightly larger than the optimal value. Failure occurs when $GCV(\alpha)$ is almost flat near the optimal alpha, leading to numerical difficulties in computing its minimum.  The \textit{cooling} procedure, as described for the L-curve, is also applied to the GCV estimation of $\alpha^{(k)}$ but with $\alpha^*$ now chosen as the current minimum of the GCV function. For clarity we summarize the steps of the inversion that are applied every iteration in Algorithm~\ref{alg1}. Here we state this for the L-curve method. The GCV solutions are found equivalently but at all steps using the L-curve instead the GCV method is applied.
\begin{algorithm}{The steps taken for the inversion at every iteration assuming the use of the L-curve to find the regularization parameter}
\label{alg1}
\end{algorithm}
\begin{enumerate}
\item Calculate the GSVD for matrix pair  $[\tilde{G}, D]$.
\item Calculate the solutions, and the associated L-curve function, for a range of $\alpha$  , the optimal $\alpha$   is found using the L-curve.
\item The cooling process is implemented for deciding whether the obtained  $\alpha$ from step 2 should be used or not.
\item With $\alpha$  from step 3, the model parameters are computed using equation \eqref{filtersoln}.
\item Density limits are implemented on model parameters, from step 4 , and then $\We$  and $\Wh$    are updated.
\item Data misfit,   $S(\bfm)$ and   $P^{\alpha}(\bfm)$ are computed for model parameters obtained from step 5.
\item If the termination criteria are satisfied the iteration terminates. Otherwise, the a priori density model is set equal to the density model from step 5 and the iteration returns to step 1.
\end{enumerate}
\section{Numerical Results: Simulated Model}\label{results}
We evaluated the use of the  L-curve and GCV in focusing inversion, as described in  Sections~\ref{Lc} and \ref{GCV}, for several synthetic  data examples. In these simulations data are calculated at $50$ stations with $10$m spacing on the surface and the inversion is required at the subsurface on a rectangular grid of $50\times 10$ with cell size $10$m, hence in this case $m=50$ and $n=500$.   The iterations are initialized with  $\bfm^{(0)}=0$, $\Wh=\We=I$ and bound constraints on the density are set such that  $0 \mathrm{gr}/\mathrm{cm}^3 \le \bfm_j \le 1 \mathrm{gr}/\mathrm{cm}^3$. The focusing parameter is fixed as $\epsilon=0.02$ for the inversions. It should be noted that the regularization parameter depends on the choice for $\epsilon$, a large value requires a larger value for $\alpha$ and generates a smoother model.  In these simulations the maximum number of iterations is set to $20$.

\begin{figure}[htb]
\begin{center}
\subfigure[Simulated Model]{\label{3a}\includegraphics[width=.45\textwidth]{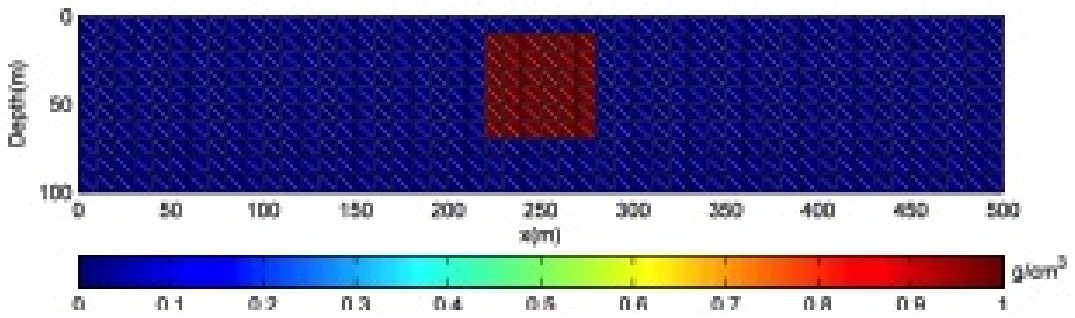}}
\subfigure[Gravity anomaly contaminated by uncorrelated noise]{\label{3b}\includegraphics[width=.45\textwidth]{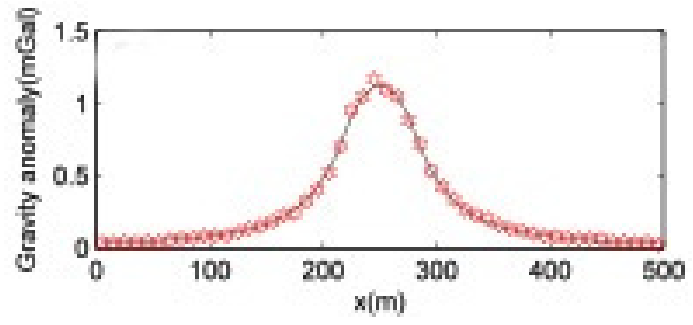}}
\caption{\label{fig3}In \ref{3a} the  synthetic model of a  body set in a grid of  square cells each of size $10$m, the density contrast of the body is $1\mathrm{gr}/\mathrm{cm}^3$. In \ref{3b} the gravity anomaly due to the synthetic model  contaminated by uncorrelated noise with $\eta_1=0.03$ and $\eta_2=.001$. The exact anomaly indicated by the solid line and the contaminated data by the symbols.}
\end{center}
\end{figure}

The synthetic gravity  data are generated in each case for the rectangular body that has density contrast equal to $1\mathrm{gr}/\mathrm{cm}^3$  with an homogeneous background, Figure~\ref{3a}. In generating noise-contaminated data we use zero mean Gaussian noise  with a standard deviation $\tilde{\sigma}_i =( \eta_1 (\bfd_{\mathrm{exact}})_i + \eta_2 \| \bfd_{\mathrm{exact}}\|)$, as indicated in Figure~\ref{3b}  for $\eta_1=.03$ and $\eta_2=.001$. In each case we calculated the   $\chi^2$ measure of the actual noise in the noisy data and report the data fidelity values, relative error $\|(\bfm_{\mathrm{exact}} - \bfm^{(K)})\|_2/\|\bfm_{\mathrm{exact}}\|_2$, and final values $\alpha^{(K)}$ for each inversion in Table~\ref{tab1}. 
\begin{figure}[htb] 
\begin{center}
\subfigure[L-curve]{\label{5a}\includegraphics[width=.45\textwidth]{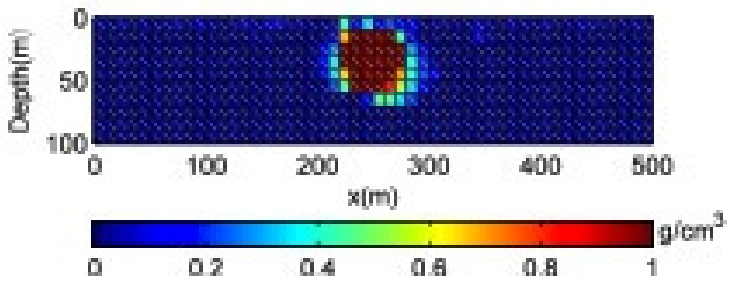}}
\subfigure[GCV]{\label{5b}\includegraphics[width=.45\textwidth]{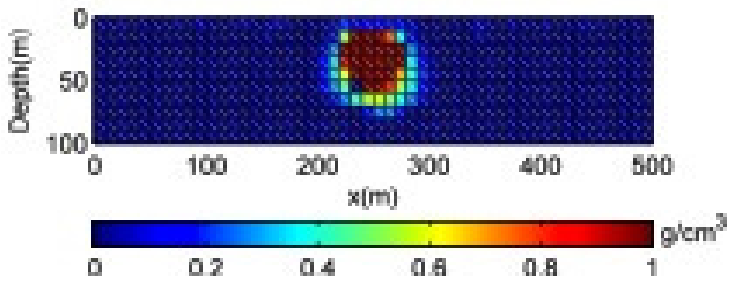}}
\caption{\label{fig5}Density model obtained from inverting the data of Figure~\ref{3a} with MS stabilizer and noise level with $\eta_1=0.03$ and $\eta_2=.001$ using bounds on density $0\mathrm{gr}/\mathrm{cm}^3\le m_j\le 1\mathrm{gr}/\mathrm{cm}^3$. The regularization parameter was found using in (a) the L-curve and in (b) the GCV.}
\end{center}
\end{figure}    
\begin{figure}[htb] 
\begin{center}
\subfigure[L-curve]{\label{6a}\includegraphics[width=.35\textwidth]{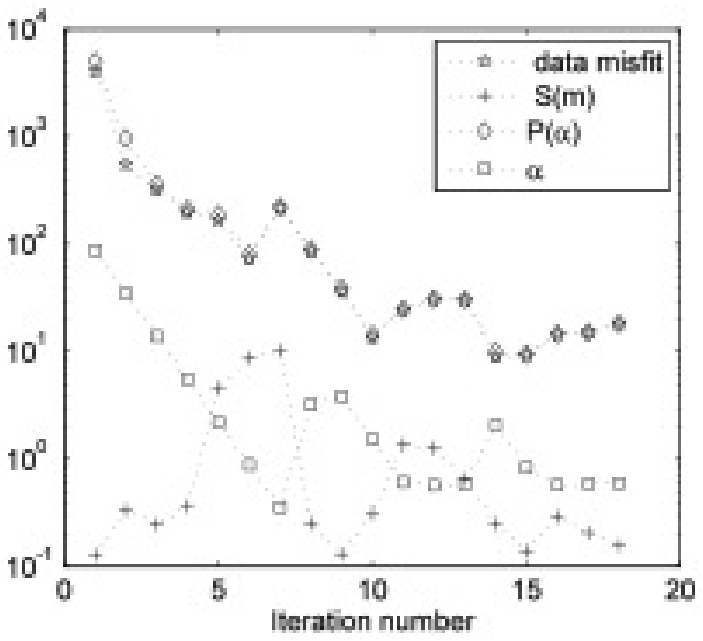}}
\subfigure[GCV]{\label{6b}\includegraphics[width=.35\textwidth]{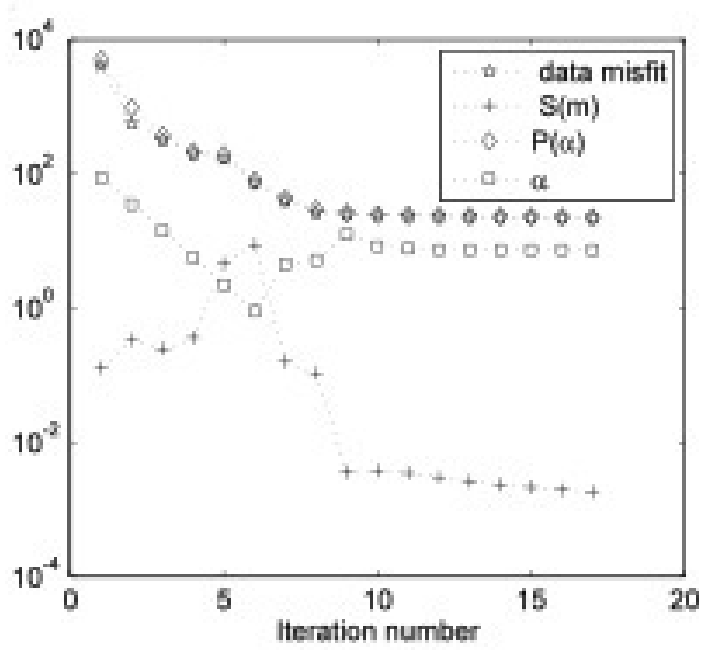}}
\caption{\label{fig6} The data fidelity $\phi(\bfd)$,  the stabilizer $S(\bfm)$, the parametric functional $P^{\alpha}(\bfm)$, and the regularization parameter,  $\alpha$ ,all plotted against iteration $k$. The regularization parameter was found using in (a) the L-curve and in (b) the GCV.}
\end{center}
\end{figure}   
   
\begin{figure}[htb]
\begin{center}
\subfigure[]{\label{5a}\includegraphics[width=.35\textwidth]{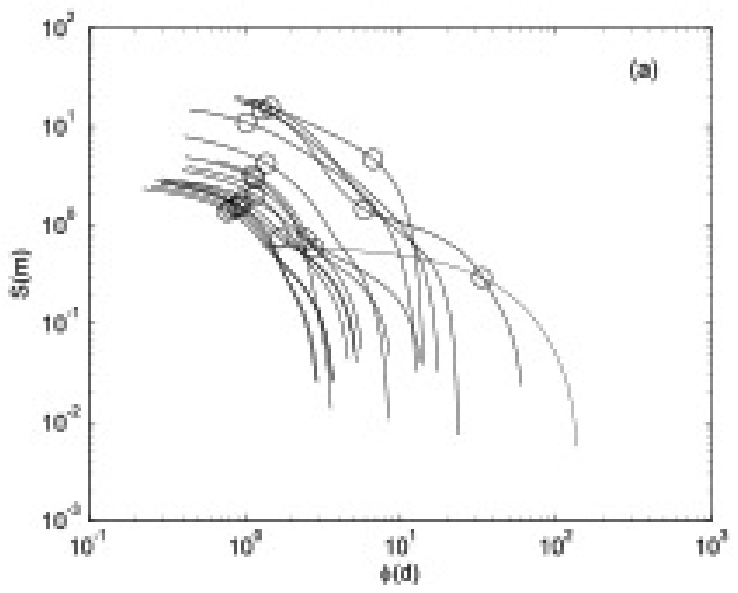}}
\subfigure[]{\label{5b}\includegraphics[width=.35\textwidth]{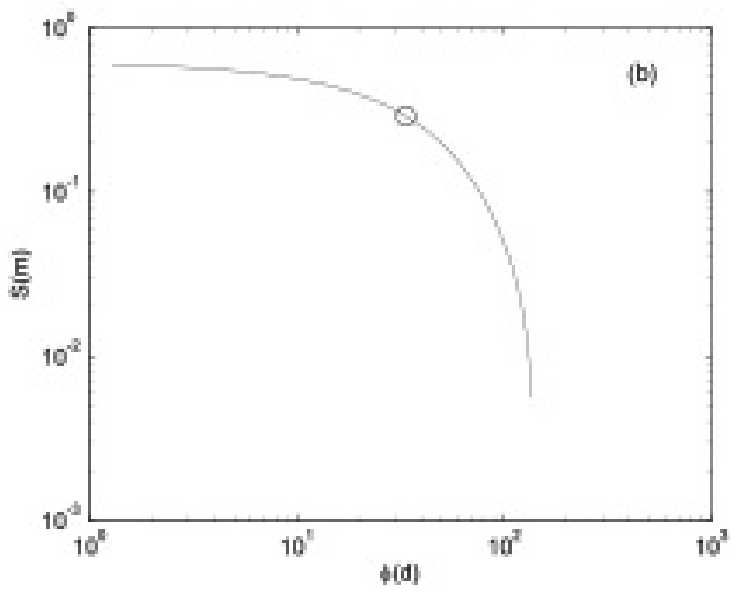}}
\subfigure[]{\label{5c}\includegraphics[width=.35\textwidth]{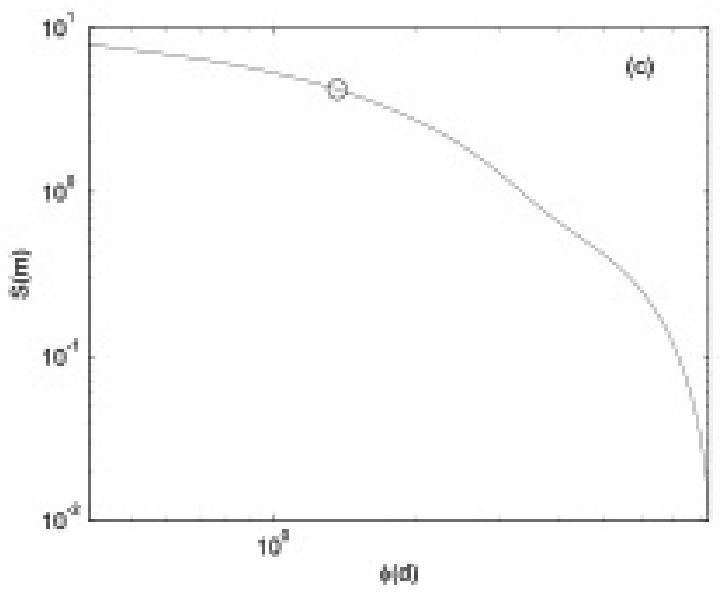}}
\subfigure[]{\label{5d}\includegraphics[width=.35\textwidth]{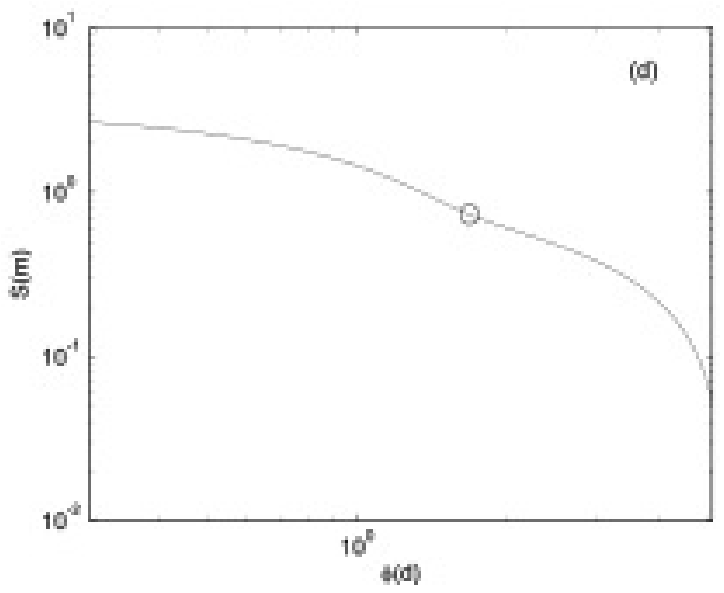}}
\caption{\label{newfig5}The L-curve for all iterations with in each case the circle showing point with maximum curvature. In figures~\ref{5b}-\ref{5d} we show the individual curves at iterations $1$, $10$ and the final stage.}
\end{center}
\end{figure}

The density models obtained for the first data set, Figure~\ref{3b} are  presented in Figures~\ref{5a} and \ref{5b} for the L-curve and GCV inversions, respectively.  In each case the geometry and density of the reconstructed models are close to those of the original model, although the inversion using the L-curve criterion is more focused. This feature was present in all the examples we have analyzed; inversion using GCV always provides a smoother reconstruction than that obtained using the L-curve.  Figures~\ref{6a}-\ref{6b} demonstrate the progression of the solutions with iteration $k$ for the data fidelity $\phi(\bfd^{(k)})$,  the stabilizer $S(\bfm^{(k)})$, the parametric functional $P^{\alpha^{(k)}}(\bfm^{(k)})$, and regularization parameter,  $\alpha^{(k)}$, again for the L-curve and GCV respectively for this first example.  In our experience the behavior indicated is consistent when using the GCV to find the regularization parameter $\alpha$; it generally decreases initially, but then increases to converge to a fixed value by the maximum number of iterations. On the other hand, for the L-curve the progression of $\alpha^{(k)}$ is more erratic, generally oscillating in final iterations toward a converged value  as shown in Figure~\ref{6a}.  Manual intervention may then be needed to force the overall convergence of the algorithm \cite{Ajo}. The main problem for the L-curve, as mentioned in section
section~\ref{Lc}, is its smooth shape, that makes it difficult to find the corner, namely the point of
maximum curvature. To illustrate we plot the L-curve for all iterations in Figure~\ref{5a}, showing
that overall the approach is successful, although at a given middle iteration the apparent corner
is missed, Figure~\ref{5c}. Still the starting and final iterations in Figure~\ref{5b} and Figure~\ref{5d}  do find
useful corners.

\begin{figure}[htb] 
\begin{center}
\subfigure[L-curve]{\label{7a}\includegraphics[width=.45\textwidth]{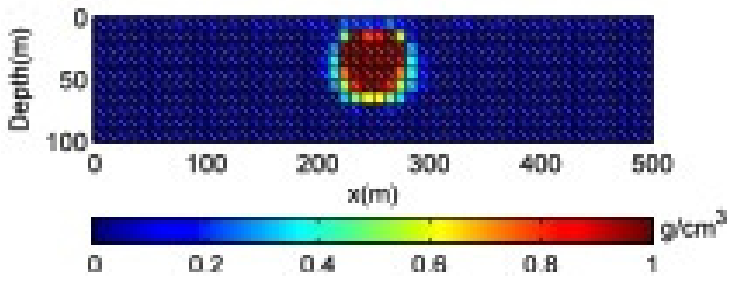}}
\subfigure[GCV]{\label{7b}\includegraphics[width=.45\textwidth]{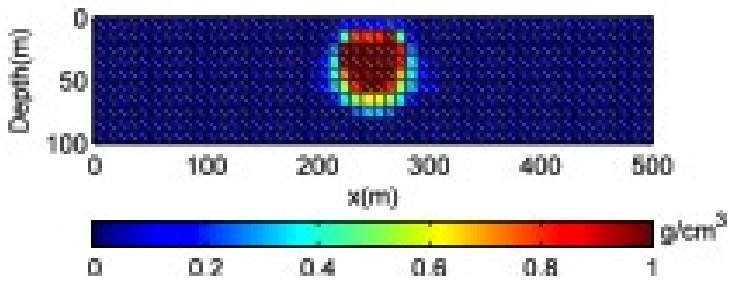}}
\caption{\label{fig7}Density model obtained from inverting the data of Figure~\ref{3b} with MS stabilizer: and noise level with $\eta_1=0.01$ and $\eta_2=.001$. The regularization parameter was found using in (a) the L-curve and in (b) the GCV.}
\end{center}
\end{figure}

\begin{figure}[htb] 
\begin{center}
\subfigure[L-curve]{\label{8a}\includegraphics[width=.45\textwidth]{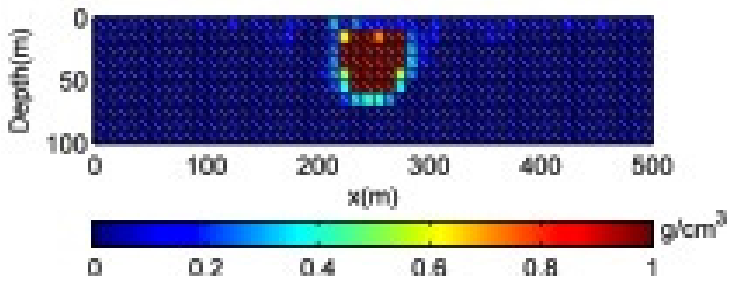}}
\subfigure[GCV]{\label{8b}\includegraphics[width=.45\textwidth]{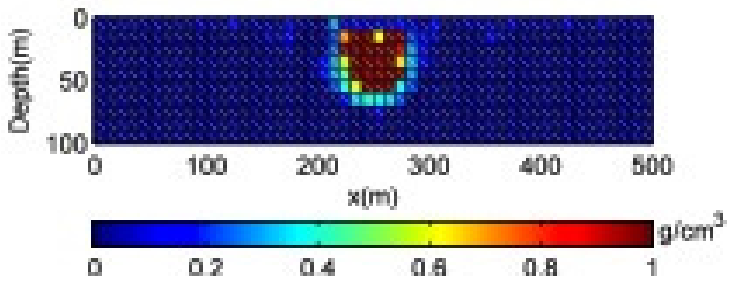}}
\caption{\label{fig8}Density model obtained from inverting the data of Figure~\ref{3b} with MS stabilizer and noise level with $\eta=0.05$ and $\eta_2=.001$. The regularization parameter was found using in (a) the L-curve and in (b) the GCV.}
\end{center}
\end{figure}

The same formulation was used to generate two further synthetic data sets, with quantitative results shown also in Table~\ref{tab1}. Noise generated using  $\eta_1=0.01$  and $\eta=0.05$ was considered, with in both cases $\eta_2=.001$. The results of the inversions are illustrated in Figures~\ref{fig7} and \ref{fig8}, respectively.   We note from Table~\ref{tab1} that the fidelity values at convergence are less than the initial $\chi^2$ measure of the noise, that always $\alpha^{(K)}_{\mathrm{\mathrm{L-curve}}} < \alpha^{(K)}_{\mathrm{GCV}}$, but that there is no fixed conclusion about the relation between the final  relative errors and fidelity estimates by   the L-curve and GCV inversions. 
We conclude that both the GCV and the L-curve are successful in providing reasonable solutions. It should be noted that increasing the focusing parameter $\epsilon$ can yield solutions which are not focused, especially as shown in Figure~\ref{fig8}. 


\begin{figure}[htb] 
\begin{center}
\subfigure[L-curve]{\label{8anew}\includegraphics[width=.45\textwidth]{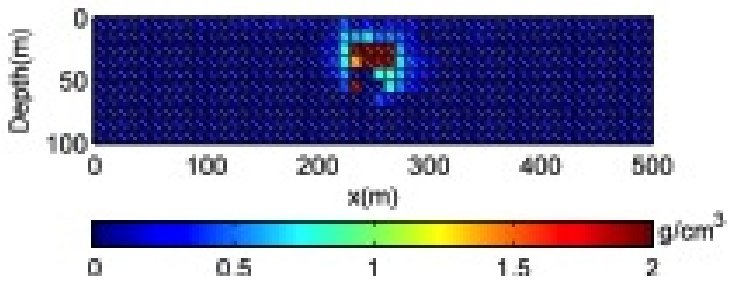}}
\subfigure[GCV]{\label{8bnew}\includegraphics[width=.45\textwidth]{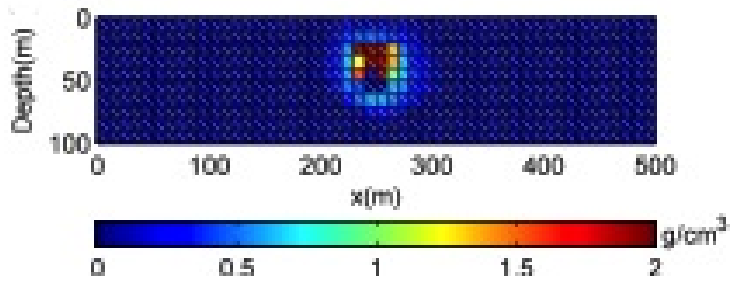}}
\caption{\label{fig8new}Density model obtained from inverting the first data set    $\eta_1=0.03$ and $\eta_2=.001$ with MS stabilizer and density limits  $ 0\mathrm{gr}/\mathrm{cm}^3\le m_j\le 2\mathrm{gr}/\mathrm{cm}^3$. The regularization parameter was found using in \ref{8anew} the L-curve; and in  \ref{8bnew} the GCV. }
\end{center}
\end{figure}

\begin{figure}[htb] 
\begin{center}
\subfigure[L-curve]{\label{9anew}\includegraphics[width=.45\textwidth]{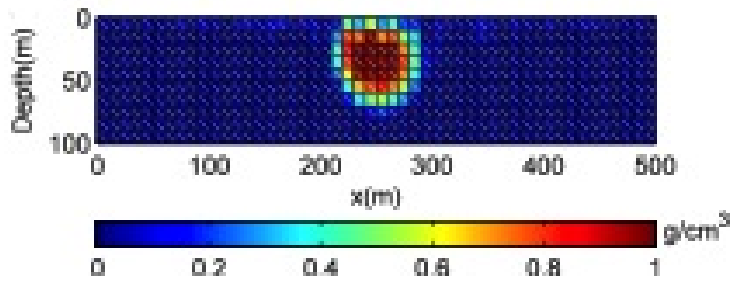}}
\subfigure[GCV]{\label{9bnew}\includegraphics[width=.45\textwidth]{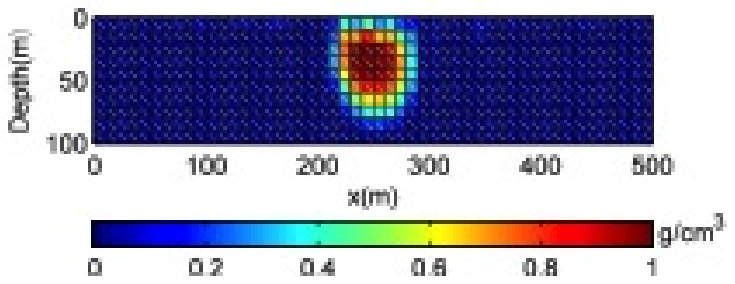}}
\subfigure[L-curve]{\label{9cnew}\includegraphics[width=.45\textwidth]{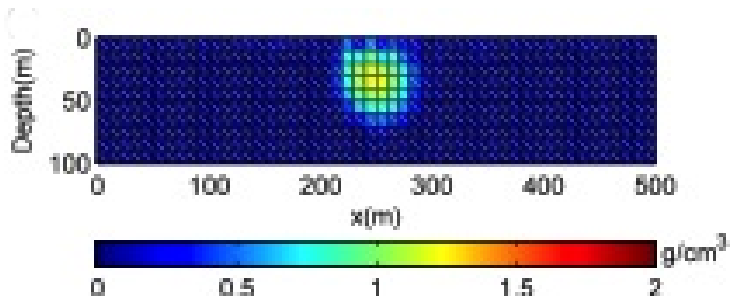}}
\subfigure[GCV]{\label{9dnew}\includegraphics[width=.45\textwidth]{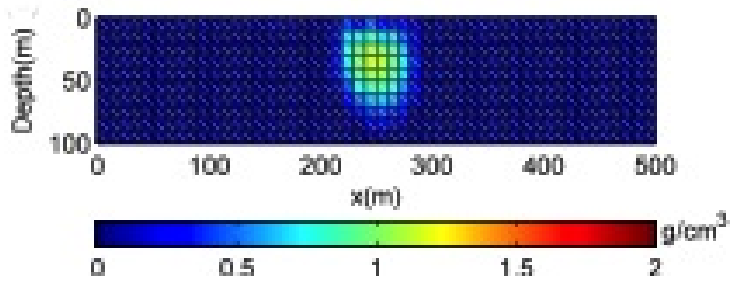}}
\caption{\label{fig9new}Density model obtained from inverting the first data set    $\eta_1=0.03$ and $\eta_2=.001$ with smoothness stabilizer. In \ref{9anew} and \ref{9bnew}  density limits  are $0\mathrm{gr}/\mathrm{cm}^3\le m_j\le 1\mathrm{gr}/\mathrm{cm}^3$. In \ref{9cnew} and \ref{9dnew}  density limits  are $0\mathrm{gr}/\mathrm{cm}^3\le m_j\le 2\mathrm{gr}/\mathrm{cm}^3$. The regularization parameter was found using in \ref{9anew} and \ref{9cnew}  the L-curve; and in  \ref{9bnew} and \ref{9dnew} the GCV.}
\end{center}
\end{figure}

\begin{table}[htb]
\begin{center}
\caption{Relative error and final values of the regularization parameter for the inverted  models.  \label{tab1}}
\vspace{0.5cm}
\begin{tabular}{|c|c|c|c|c|c|c|c|} \hline
Figure & $\chi^2$& \multicolumn{2}{|c|}{$\alpha^{(K)}$}&\multicolumn{2}{|c}{Relative Error}&\multicolumn{2}{|c|}{Fidelity $ \phi(\bfd)$}\\  \hline
 & measure& L-curve & GCV& L-curve& GCV & L-curve& GCV \\ \hline
 $3$&$51.23$&$0.62$&$7.41$&$0.4270$& $0.4025$&$19.14$&$21.93$\\ \hline
$6$&$47.26$&$0.87$&$14.15$&$0.3376$& $0.3506$&$26.27$&$24.96$\\ \hline
$7$&$40.03$&$0.50$&$6.85$&$0.4014$&$0.4013$ &$21.63$&$13.61$\\ \hline
$8$&$51.23$&$0.63$&$8.20$&$0.7708$&$0.6910$ &$17.21$&$20.80$\\ \hline
$9$a-b &$51.23$&$27.04$&$429.38$&$0.4264$&$0.4026$ &$19.64$&$16.77$\\ \hline
$9$c-d&$51.23$&$27.23$&$441.86$&$0.4404$&$0.4069$ &$17.95$&$16.10$\\ \hline
\end{tabular}
\end{center}
\end{table}
To assess both the impact of the choice of the bounds on the convergence properties for the solution and choice of the regularization parameter  $\alpha$, we investigated two additional situations. First we implemented the same problem as given in figure~\ref{fig3}, with noise $\eta_1=0.03$ and $\eta_2=0.001$, but inverted now with upper bounds on the density changed to $2\mathrm{gr}/\mathrm{cm}^3$. The results are illustrated in figure~\ref{fig8new} and detailed as before in Table~\ref{tab1}. We see that the solutions are more focused, as anticipated from the previous work of \cite{Port} but the relative error overall is increased and the fidelity of the solution is also decreased. On the other hand, it is of greater interest for the purposes of this study to observe that the performance of the parameter choice techniques is independent of the upper bound, and the parameters are found stably independent of the constraint bounds. To determine the necessity of using the MS instead of a smoothness stabilizer, we also considered the results obtained using smoothness stabilizer, i.e. the $\We$  in \eqref{model}  replaced with the approximation for the second derivative of the model parameters, for the situations in figures~\ref{fig5} and \ref{fig8new}. These results are also detailed in Table~\ref{tab1}  and illustrated in figures~\ref{9anew}-\ref{9bnew} and \ref{9cnew}-\ref{9dnew},  respectively. They demonstrate the relative insensitivity to density limits of the smoothness-stabilizer obtained solutions. On the other hand, the solutions lack the contrast that is achieved using the MS regularization. Overall, the solutions obtained with GCV are apparently more robust than those with the L-curve. It should be noted that using a non-$\ell_2$   measure of the derivative in geophysical inversion leads to a strongly piecewice constant, or blocky, reconstruction with sharp jumps, see e.g. \cite{Far:98}.   

\section{Numerical Results: Practical Data}\label{realresults}
\subsection{Geological Context}
The data used for inversion was acquired over the Safo mining camp in Maku-Iran which is well known for manganese ores. Geologically this area  is located in the Khoy ophiolite zone, in the northwest of Iran. Some manganese and iron-manganese deposits are found within sedimentary pelagic rocks and radiolarian cherts which are accompanied by Khoy ophiolite, \cite{Imama}. Most of these deposits  have little reserve;   the Safo deposit is the only viable  area distinguished  so far for mining. In the Safo deposit, depositions of manganese have been found to occur in different horizons within pelagic rocks. Mineralogically, pyrolusite, bixibite, braunite and hematite are the main minerals present in ore, of which the pyrolusite is the dominant ore \cite{Imama}, and Calcite with quartz and barite present as minor phases. The banded, massive and disseminated textures are seen in orebodies. Manganese content varies from   $7.4\%$ to  $69.1\%$  in different regions of the area \cite{Imama}.

\subsection{Gravity anomaly}

\begin{figure}[htb] 
\begin{center}
\subfigure[L-curve]{\label{9a}\includegraphics[width=.45\textwidth]{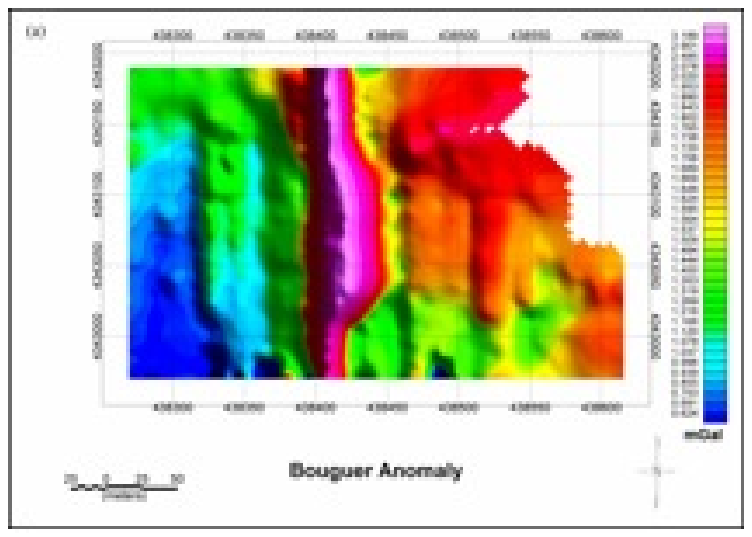}}
\subfigure[GCV]{\label{9b}\includegraphics[width=.45\textwidth]{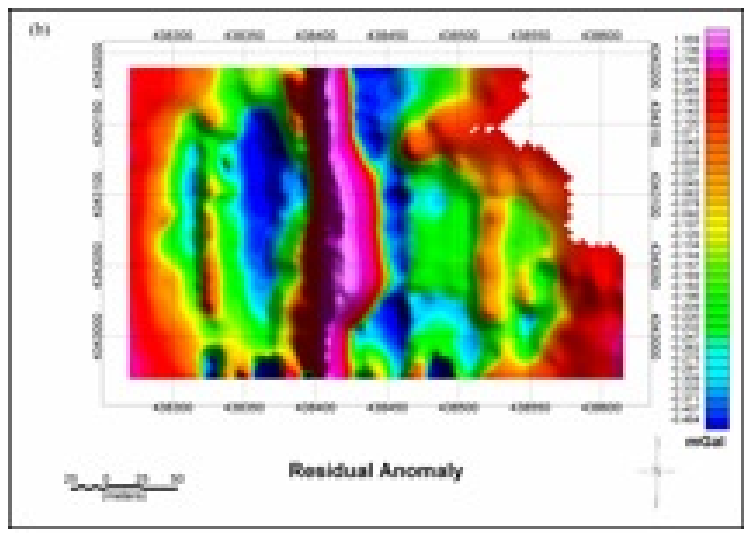}}
\caption{\label{fig9} Bouguer anomaly over the Safo manganese mine in \ref{9a} and the residual anomaly over the Safo manganese mine in \ref{9b}.}
\end{center}
\end{figure}

\begin{figure}[htb] 
\begin{center}
\includegraphics[width=.45\textwidth]{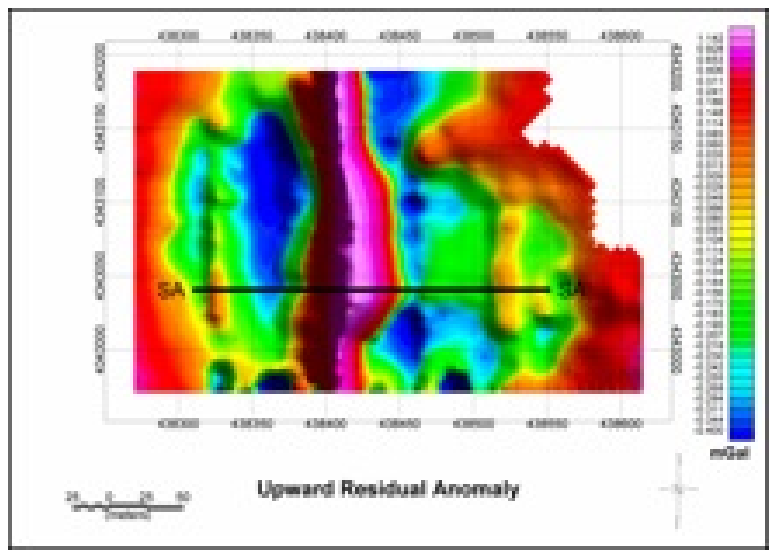}
\caption{\label{fig10} Upward continuation of the residual anomaly to height $2.5$m.}
\end{center}
\end{figure}

The area of the gravity survey extends between UTM coordinates $[438276$ $438609]$ west and $[4342971$ $4343187]$ north, Z38. The gravity survey was performed by the gravity branch of the institute of Geophysics, University of Tehran. The measurements were corrected for effects caused by instruments and tidal drift, latitude, free air and the Bouguer correction to yield the Bouguer gravity anomaly, Figure~\ref{9a}. The Bouguer anomaly displays extreme magnitudes in the central of the area in the north-south direction, related to mineral occurrence which has  high density contrast with the host rocks. This geologic structure is therefore clearly suitable  for using a 2-D algorithm. The residual anomaly was obtained by subtracting the regional anomaly from the Bouguer anomaly using a polynomial fitting method, Figure~\ref{9b}. One of the recommended steps in potential field inversion is upward continuation of data to a height of half the thickness of the shallowest cell which removes near surface effects without noticeably degrading the data. Figure~\ref{fig10} shows upward continuation of the residual data up to $2.5$m. 

\subsection{Inversion result}
\begin{figure}[htb] 
\begin{center}
\subfigure[L-curve]{\label{11a}\includegraphics[width=.45\textwidth]{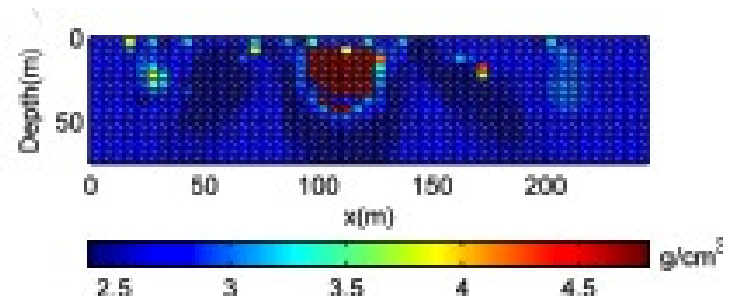}}
\subfigure[GCV]{\label{11b}\includegraphics[width=.45\textwidth]{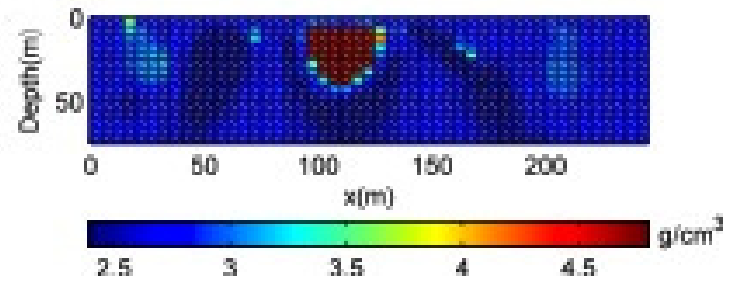}}
\caption{\label{fig11} The density models obtained by inverting field gravity data (profile SA). The regularization parameter was found using in (a) the L-curve and in (b) the GCV.}
\end{center}
\end{figure}

\begin{figure}[htb] 
\begin{center}
\subfigure[L-curve]{\label{13anew}\includegraphics[width=.45\textwidth]{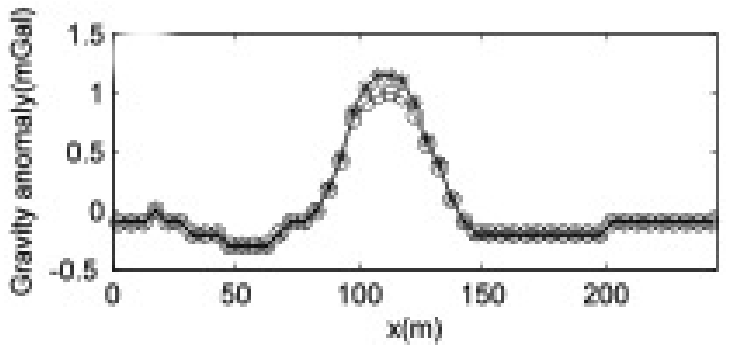}}
\subfigure[GCV]{\label{13bnew}\includegraphics[width=.45\textwidth]{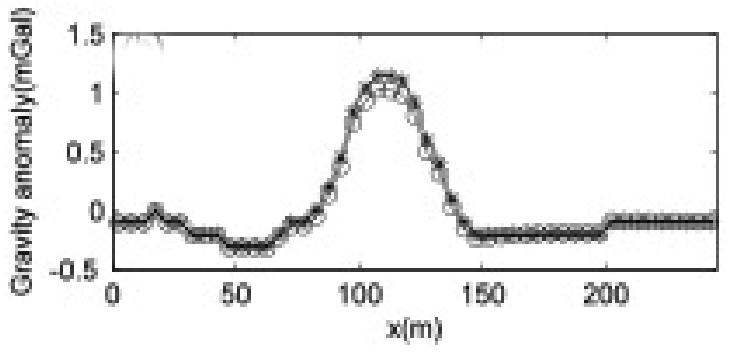}}
\caption{\label{fig13} The field gravity data (stars) and computed data for the reconstructed density model (circles). The regularization parameter was found using in \ref{13anew} the L-curve; in \ref{13bnew}  the GCV.}
\end{center}
\end{figure}

\begin{figure}[htb] 
\begin{center}
\subfigure[L-curve]{\label{12a}\includegraphics[width=.35\textwidth]{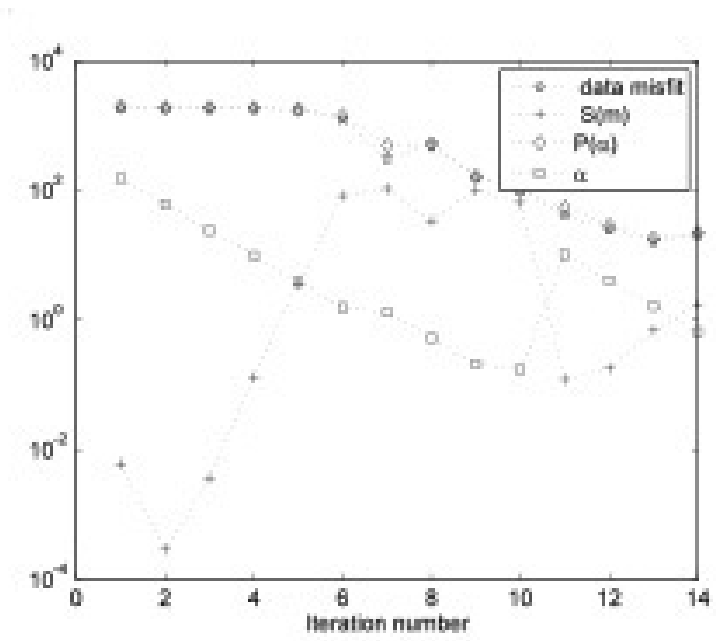}}
\subfigure[GCV]{\label{12b}\includegraphics[width=.35\textwidth]{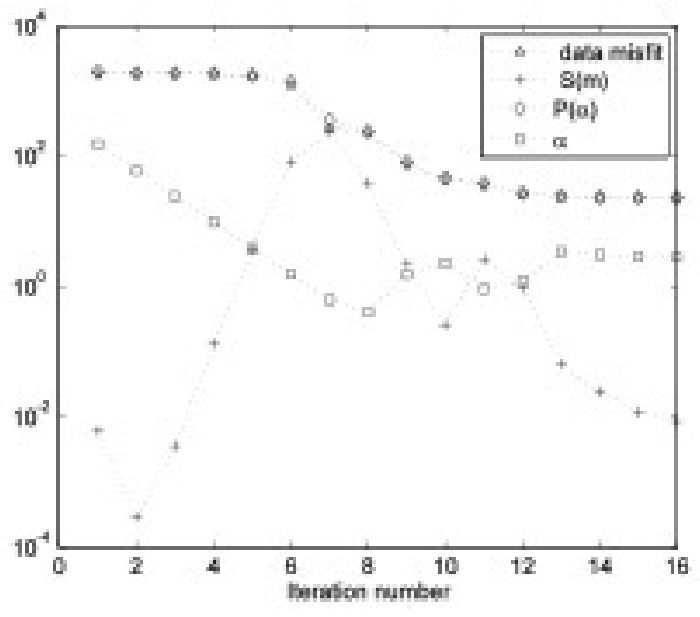}}
\caption{\label{fig12} The data fidelity $\phi(\bfd)$,  the stabilizer $S(\bfm)$, the parametric functional $P^{\alpha}(\bfm)$, and the regularization parameter,  $\alpha$ ,all plotted against iteration $k$.  The regularization parameter was found using in (a) the L-curve and in (b) the GCV.}
\end{center}
\end{figure}

A profile of the anomaly (SA) consisting of $49$ data measurements,  sampled every $5$m, is chosen for inversion. 
The subsurface is divided into $49\times 15$  square cells of size $5$m, hence in this case $m=49$ and $n=735$. Based on geological information \cite{Imama}, the background density is set to $2.8\mathrm{gr}/\mathrm{cm}^3$ and the density limits for the inversion are $2.4 \mathrm{gr}/\mathrm{cm}^3 \le \bfm_j \le 4.7 \mathrm{gr}/\mathrm{cm}^3$. The maximum number of iterations was set to $20$. Each datum is assigned a Gaussian error as in the simulated cases, here with $\eta_1=.05$ and $\eta_2=.001$.  Figures~\ref{fig11}a-b illustrate the reconstructed density model from the inversion of profile SA using the L-curve and GCV  methods for estimating the regularization parameter, yielding   $\alpha^{(K)}_{\mathrm{L-curve}}=0.65$ and $\alpha^{(K)}_{\mathrm{GCV}}=4.87$ respectively. 
Figures~\ref{13anew}-\ref{13bnew}   illustrate the profile of the anomaly (SA) which is used for the inversion, indicated by the stars, and the resulting values obtained from reconstructed models in figures~\ref{11a}-\ref{11b}, denoted in each case by the circles. 
Both solutions clearly represent the density contrast and geometry for the occurrence of manganese ore, figures~\ref{11a}-\ref{11b}. The horizontal extension of the obtained model is about $30$m and the vertical extension shows a depth interval approximately between $5$m and  $35$m. 
These results are close to those obtained by Borehole
drilling on the site; which show extension of manganese ores from $3-4$m to $25-30$m in the
subsurface along the north-south direction \cite{Noori}. The data fidelity, the stabilizer, the parametric
functional and regularization parameter, with iteration $k$ , are shown in  Figures~\ref{12a}-\ref{12b}. The convergence histories have properties for the practical data that are similar to those for the simulated data sets.

\section{Conclusions}\label{sec:conc}
Tikhonov regularization with the minimum support stabilizer has been demonstrated to yield non-smooth solutions and is thus an appropriate approach for recovery of  geological structures with sharp boundaries. The presented algorithm is flexible and allows variable weighting in the stabilizer, including  depth weighting, a priori density ranges for the domain and the inclusion of hard constraints for the a priori information in the  inversion process. The L-curve criterion and GCV method for estimating the regularization parameter were discussed, and characteristics of each of them for obtaining a solution were introduced. Numerical tests using synthetic data have demonstrated feasibility of applying both methods in the iteratively reweighted algorithm. The regularization parameter is seen to converge as the number of iterations increases. Although GCV leads to inverse solutions which are slightly smoother than those obtained by  the L-curve, both  recovered models are close to the original model. For this  small-scale  problem, it is shown that the GSVD can be used in the algorithm, demonstrating the filtering of the solution.  Moreover, this use of the GSVD, which might generally be assumed to be too expensive, is beneficial and worthwhile in the context of regularization parameter estimation as shown here. For large-scale problems it is anticipated that a randomized GSVD needs to be developed, along the lines of the randomized SVD that was introduced in \cite{Liberty}.  Finally the method was used on a profile of gravity data acquired over the Safo manganese mine in the northwest of Iran. The result shows a density distribution in subsurface from about $5$m to $35$m in depth and about $30$m  horizontally.  Future work will consider the inclusion of statistical weighting in the solution and the use of regularization parameter estimation using statistical approaches, \cite{mere:09}.
\section{Acknowledgements}  Rosemary Renaut acknowledges the support of AFOSR grant 025717: ``Development and Analysis of Non-Classical Numerical Approximation Methods", and 
NSF grant  DMS 1216559:   ``Novel Numerical Approximation Techniques for Non-Standard Sampling Regimes". We would also like to thank the two anonymous referees who raised interesting questions that lead us to include additional results and clarifications of the method.

\end{document}